\documentclass[11pt]{article}

\usepackage[T1]{fontenc}
\usepackage[utf8]{inputenc}
\usepackage[american]{babel}
\usepackage{amsmath,amssymb,amsfonts}
\usepackage{mathtools}
\usepackage{amsthm}
\usepackage{microtype}
\usepackage{geometry}
\usepackage{hyperref}

\hypersetup{
  colorlinks=true,
  linkcolor=blue,
  citecolor=blue,
  urlcolor=blue
}

\newcommand{\lc}[2]{\left\{^{\,#1}_{#2}\right\}}
\newcommand{\dd}{\mathrm{d}}
\newcommand{\sginv}{{}^{\mathrm S}g}

\theoremstyle{plain}
\newtheorem{theorem}{Theorem}
\newtheorem{lemma}[theorem]{Lemma}
\newtheorem{proposition}[theorem]{Proposition}
\newtheorem{corollary}[theorem]{Corollary}

\theoremstyle{definition}
\newtheorem{definition}[theorem]{Definition}
\newtheorem{assumption}[theorem]{Assumption}

\theoremstyle{remark}
\newtheorem{remark}[theorem]{Remark}

\title{Gravitational and Maxwell--Proca-Type Structures in a Variational\\
Metric-Affine Framework}

\author{Oldřich Stolín\\[0.5em]
\small Mathematical Institute in Opava, Silesian University in Opava\\
\small Na Rybníčku 626/1, 746 01 Opava, Czech Republic\\
\small \texttt{oldrich.stolin@math.slu.cz}}

\date{}

\begin{document}

\maketitle

\begin{abstract}
We study Einstein's nonsymmetric metric-affine Lagrangian density with
a generally nonsymmetric metric and an affine connection treated as
independent variational variables. The resulting connection equation
belongs to the classical Einstein--Hlavat{\'y}--Tonnelat framework, while
the unrestricted metric Euler--Lagrange equation is analyzed without
a divergence constraint on the antisymmetric metric density.

Its symmetric part takes the form of Einstein-type gravitational
equations with geometrically induced source terms and reduces to the
vacuum Einstein equations in the torsion-free limit. The antisymmetric
part becomes a geometric constitutive relation based on the
irreducible decomposition of torsion. Its contracted four-divergence
yields a general vector-field equation with geometric source and
response terms. Under a trace--axial reduction, \(Q=0\)
gives a source-free Maxwell-type equation, whereas constant nonzero
\(Q\) gives a Proca-type equation.

Independently, a trace of the generalized Bianchi identities yields the
homogeneous Maxwell-type pair associated with the torsion trace, while
the connection equation implies a Lorenz-type condition. Thus gravitational, constitutive, Maxwell--Proca-type, and Lorenz-type
structures arise as compatible consequences of one metric-affine
variational system. The antisymmetric dynamical level is generated by
the contracted four-divergence of its constitutive level. No independent
electromagnetic or matter field is introduced.
\end{abstract}



\medskip

\section{Introduction}

\paragraph{Classical background and variational setting.}
The geometric description of gravitation naturally raises the question
whether other classical field structures can arise from the same
geometric and variational framework rather than being introduced as
independent fields. This question was central to Einstein's later
unified-field program. Nonsymmetric metric-affine geometry provides a
natural setting for such an approach: the metric and the affine
connection contain additional components beyond those of general
relativity, while torsion and distinct curvature contractions provide
geometric objects with vector-field and constitutive-type structures.

The historical development of Einstein's nonsymmetric metric-affine
approach and its relation to other unified-field programs are reviewed,
for example, by Goenner \cite{goenner2004,goenner2014}. Primary versions
of the nonsymmetric program relevant here were developed by Einstein and
Einstein--Straus \cite{einstein1945_generalization,einstein_straus1946}.
The corresponding connection problem was studied systematically by
Hlavat{\'y} and, independently in a different form, by Tonnelat
\cite{hlavaty1957,tonnelat1966}. Related affine-geometric approaches to
gravitational and electromagnetic structures were developed, among
others, by Ferraris and Kijowski \cite{ferraris_kijowski1982}.

Several later constructions obtained electromagnetic- or vector-field
structures from nonsymmetric or torsion-based geometries by different
mechanisms. McKellar derived the Einstein--Maxwell equations from an
asymmetric-connection variational principle; Moffat and Ragusa obtained
Proca- or Maxwell-type antisymmetric fields in linearized nonsymmetric
models; and Légaré and Moffat analyzed first-order field equations,
generalized Bianchi identities, and conservation laws
\cite{mckellar1979,moffat1995_new,ragusa2000,legare_moffat1995}.
Analyses of related nonsymmetric metric theories also exposed delicate
consistency and ghost questions \cite{damour_deser_mccarthy1993}.
Recent work continues the study of differential identities in Einstein nonsymmetric geometry \cite{wanas_nabil_abdelhamid_elabd2025} and of geometric gravity--electromagnetism unification, including contributions in this journal \cite{wanas_nabil_sherif2026,yaremenko2026}.

In the present paper we return to the Lagrangian density used in
Einstein's nonsymmetric metric-affine theory. The fundamental
variational variables are a generally nonsymmetric covariant tensor
field \(g\) and an affine connection \(\nabla\), represented in a local
coordinate frame by \(g_{ij}\) and \(\Gamma^i{}_{jk}\), respectively.
They are varied independently, and no additional matter or
electromagnetic field is introduced. The construction is generally
covariant, and the Lagrangian density is invariant under the discrete
pseudo-Hermitian transposition introduced below. The resulting
connection Euler--Lagrange equation is the classical
Einstein--Hlavat{\'y}--Tonnelat equation and determines the affine
connection from the nonsymmetric metric under the corresponding
regularity conditions. We use this classical solution of the connection
problem without reproducing its complete algebraic classification.

The variational distinction concerns the admissible metric variations.
Einstein's constrained prescription imposes
\(\partial_j\mathfrak g^{[ij]}=0\) prior to variation, where
\(\mathfrak g^{ij}:=\omega g^{ij}\) denotes the contravariant metric density
introduced below;
the admissible variations must therefore satisfy the corresponding
linearized condition. See the historical analysis in
Ref.~\cite[Sec.~9.3]{goenner2014}. We impose neither restriction.
This comparison is with that constrained prescription, not with every
formulation of Einstein's nonsymmetric theory. The unrestricted metric
Euler--Lagrange equation must therefore be analyzed on its own. Its
antisymmetric part supplies a constitutive relation whose contracted
four-divergence yields the vector-field equation, whereas the homogeneous
Maxwell-type identity follows from the second curvature trace.

\paragraph{Main results.}
The symmetric part can be written as an Einstein-type gravitational
equation for the symmetric metric \(g_{(ij)}\). Its right-hand side is
formed entirely from the non-Riemannian components of the same
metric-affine geometry. In the torsion-free reduction, the connection
becomes the Levi--Civita connection of \(g_{(ij)}\), the induced
geometric source vanishes, and the equation reduces to the vacuum
Einstein equation.

The antisymmetric part has a different structure. Using the irreducible
decomposition of torsion into its trace, axial, and traceless tensor
components with respect to the proper orthochronous component of the
local Lorentz group associated with \(g_{(ij)}\), it can be written as
a geometric constitutive relation in which the irreducible torsion
components acquire distinct algebraic and differential roles. The
torsion trace supplies the vector-potential- and field-strength-type
contribution, whereas the remaining torsion components and the
non-Riemannian symmetric part of the connection produce response and
source-like terms. Thus the antisymmetric Euler--Lagrange equation is
not identified directly with a Maxwell equation; rather, its contracted
four-divergence gives the corresponding vector-field equation.
Accordingly, the constitutive and dynamical levels are not introduced
independently: the antisymmetric metric Euler--Lagrange equation
provides the constitutive level, while its contracted four-divergence
generates the corresponding dynamical level.

This divergence equation contains two distinguished limits. Under the
trace--axial reduction considered below, \(Q=0\) gives a source-free
Maxwell-type equation, while \(Q=\mathrm{const}\neq0\) gives a
Proca-type equation with a geometrically generated potential term.
Independently of this constitutive equation, the second curvature trace
is related to the exterior derivative of the torsion trace, and its
cyclic identity gives the homogeneous Maxwell-type pair. The connection
Euler--Lagrange equation also implies a Lorenz-type condition for the
torsion-trace density. The homogeneous equations, the constitutive
relation, and its four-divergence therefore arise from distinct but
mutually compatible parts of the same metric-affine structure.

The symmetric and antisymmetric parts of the same metric
Euler--Lagrange equation therefore generate parallel but differently
ordered differential hierarchies. In the symmetric part, the field
equation is already of Einstein type and its covariant divergence gives
a conservation law. In the antisymmetric part, the metric equation
first assumes a constitutive form; its contracted divergence produces
the vector-field equation, and one further divergence gives the
corresponding conservation identity. The antisymmetric hierarchy is
therefore shifted by one differential level relative to the symmetric
gravitational hierarchy.

\section{Geometric Framework and Notation}
\label{sec:geom}

We first establish the geometric framework and notation used throughout
this paper. We specify the independent metric-affine variables and their
regularity assumptions, introduce the torsion and covariant-derivative
conventions, and recall the irreducible decomposition of torsion with
respect to the Lorentzian metric defined by $g_{(ij)}$. We also define
the pseudo-Hermitian transposition employed in the subsequent variational
analysis.

\subsection{Basic variables and regularity}

Let \(M\) be a four-dimensional differentiable manifold equipped with a
generally nonsymmetric covariant tensor field \(g\) and an affine
connection \(\nabla\) that is not assumed to be torsion-free. In an
arbitrary but fixed local coordinate frame, their components are denoted
by \(g_{ij}\) and \(\Gamma^i{}_{jk}\), respectively.

\begin{assumption}[Metric-affine variables]
The tensor field \(g\) and the affine connection \(\nabla\) are treated
as independent variational variables. Equivalently, in a fixed local
coordinate frame, their components \(g_{ij}\) and
\(\Gamma^i{}_{jk}\) are varied independently. No symmetry condition is
imposed on \(g\), and \(\nabla\) is not assumed to be torsion-free.
\end{assumption}

The symmetric part of \(g\), whose components are \(g_{(ij)}\), defines
the Lorentzian metric structure, whereas \(g_{[ij]}\) are the components
of its antisymmetric part.

\begin{assumption}[Regularity and signature]
\label{ass:regularity_signature}
Throughout the paper,
\[
g<0,
\qquad
{}^{\mathrm S}g<0,
\qquad
g_{(ij)}
\text{ has signature }
(+,+,+,-),
\]
where
\[
g:=\det(g_{ij}),
\qquad
{}^{\mathrm S}g:=\det(g_{(ij)}).
\]
\end{assumption}

The matrix \((g_{ij})\) is invertible, and its inverse is denoted by
\((g^{ij})\), so that
\[
g_{ij}g^{jk}
=
g_{ji}g^{kj}
=
\delta_i^k.
\]
The inverse matrix of \((g_{(ij)})\) is denoted by
\((\sginv^{ij})\), so that
\[
g_{(ij)}\sginv^{jk}
=
g_{(ji)}\sginv^{kj}
=
\delta_i^k.
\]
Unless stated otherwise, tensor indices are raised and lowered with the
symmetric metric and its inverse, whose components are \(g_{(ij)}\) and
\(\sginv^{ij}\). Thus \(g^{ij}\) denotes the components of the inverse
of the full nonsymmetric tensor \(g\) and should not be confused with
the tensor obtained by raising the indices of \(g_{ij}\) with
\(\sginv^{ij}\).

We use the two volume densities
\[
\omega:=\sqrt{-g},\qquad
\omega_{\mathrm S}:=\sqrt{-{}^{\mathrm S}g},\qquad
\mathfrak g^{ij}:=\omega g^{ij}.
\]
The full density \(\omega\) is the one entering the variational
quantities. The symmetric-metric density \(\omega_{\mathrm S}\) is used
only when the Levi--Civita divergence is written in volume form. In
particular, for an antisymmetric contravariant tensor \(X^{ik}\),
\begin{equation}
\label{eq:LC_divergence_volume_form}
X^{ik}{}_{;\,k}
=
\frac{1}{\omega_{\mathrm S}}
\partial_k\!\left(\omega_{\mathrm S}X^{ik}\right).
\end{equation}
Thus no second family of tensor-density symbols is required below.

\subsection{Torsion and covariant derivatives}

In the chosen local coordinate frame, the coefficients of the affine
connection decompose as
\[
\Gamma^i{}_{jk}
=
\Gamma^i{}_{(jk)}
+
\Gamma^i{}_{[jk]}.
\]

\begin{definition}[Torsion and its trace]
\label{def:torsion}
With the convention used throughout this paper, the torsion tensor
\(S\) has the components
\begin{equation}
\label{eq:torsion_definition}
S^i{}_{jk}
:=
\Gamma^i{}_{[jk]}.
\end{equation}
Its trace one-form has the components
\begin{equation}
\label{eq:torsion_trace_potential}
S_i
:=
S^j{}_{ij}.
\end{equation}
\end{definition}

Our convention differs by a factor \(2\) from the frequently used
definition
\[
T^i{}_{jk}:=\Gamma^i{}_{jk}-\Gamma^i{}_{kj}
=2S^i{}_{jk}.
\]

Because the affine connection is not assumed to be symmetric in its
lower indices, their order must be distinguished. For a covector field
\(A\), with components \(A_i\), we use
\begin{align}
\label{eq:covector_derivatives}
A_{\substack{i\\+}\,\substack{;\,k\\\vphantom{+}}}
&:=
A_{i,k}
-
A_s\Gamma^s{}_{ik},
&
A_{\substack{i\\-}\,\substack{;\,k\\\vphantom{-}}}
&:=
A_{i,k}
-
A_s\Gamma^s{}_{ki},
\nonumber\\
A_{\substack{i\\0}\,\substack{;\,k\\\vphantom{0}}}
&:=
A_{i,k}
-
A_s\Gamma^s{}_{(ik)},
&
A_{i;\,k}
&:=
A_{i,k}
-
A_s\lc{s}{ik}.
\end{align}
The decorations \(+\), \(-\), and \(0\) refer respectively to the
connection coefficients
\(\Gamma^i{}_{jk}\), \(\Gamma^i{}_{kj}\), and
\(\Gamma^i{}_{(jk)}\), whereas an undecorated semicolon denotes the
Levi--Civita covariant derivative associated with the symmetric metric.
In particular, the symbol \(0\) labels the symmetrized affine
connection and is unrelated to the weight of a tensor density.

The same convention applies to arbitrary tensor fields, with the usual
opposite signs of the connection terms for covariant and contravariant
indices. In mixed decorated derivatives, each tensor index is
differentiated according to its own decoration.

All tensor densities used below have weight one. For a scalar density
\(\rho\) and a contravariant vector density \(\mathcal V^i\), we define
\begin{align}
\rho_{\substack{;\,k\\0}}
&:=
\rho_{,k}
-
\rho\,\Gamma^s{}_{(sk)},
\label{eq:scalar_density_derivative_zero}
\\
\mathcal V^i{}_{\substack{;\,k\\0}}
&:=
\mathcal V^i{}_{,k}
+
\Gamma^i{}_{(sk)}\mathcal V^s
-
\Gamma^s{}_{(sk)}\mathcal V^i.
\label{eq:vector_density_derivative_zero}
\end{align}
The corresponding Levi--Civita derivatives are obtained by replacing
\(\Gamma^i{}_{(jk)}\) with \(\lc{i}{jk}\). Mixed decorated density derivatives below use the correspondingly
ordered connection on tensor indices and the symmetrized trace
\(\Gamma^s{}_{(sk)}\) in the density-weight term. In particular,
\(\omega_{\mathrm S;k}=0\), whereas
\(\omega_{\substack{;\,k\\0}}=0\) follows from the connection equation,
Eq.~\eqref{eq:omega_covariant_derivative_zero}. Neither statement
allows the volume density to change silently when the connection changes.
The volume covector in \eqref{eq:Aplus} is defined separately by its
explicit ordered trace.

We also introduce the difference between the Levi--Civita derivative
and the covariant derivative defined by the symmetrized affine
connection. For a tensor field \(T\) of type \((r,s)\), we define
\begin{equation}
\label{eq:Delta_derivative_definition}
\left(
T^{i_1\ldots i_r}{}_{j_1\ldots j_s}
\right)_{\substack{;\,k\\\Delta}}
:=
\left(
T^{i_1\ldots i_r}{}_{j_1\ldots j_s}
\right)_{;\,k}
-
\left(
T^{i_1\ldots i_r}{}_{j_1\ldots j_s}
\right)_{\substack{;\,k\\0}}.
\end{equation}
The decoration \(\Delta\) isolates the contribution arising from the difference between the Levi--Civita covariant derivative and the covariant derivative defined by the symmetrized affine connection. This separation will be useful in the subsequent geometric decompositions and interpretations.

\subsection{Irreducible decomposition of torsion}

We use the standard four-dimensional decomposition of torsion into its
trace-vector, axial-vector, and traceless tensor parts with respect to
the proper orthochronous component of the local Lorentz group associated 
with the symmetric metric \(g_{(ij)}\); see
\cite[Appendix~B]{hehl_mccrea_mielke_neeman1995},
\cite[Sec.~2.1]{shapiro2002}, and
\cite[Sec.~2.2]{capozziello_stornaiolo2007}.
The numerical factors below correspond to our convention
\(S^i{}_{jk}=\Gamma^i{}_{[jk]}\).

First, the trace \(S_i=S^k{}_{ik}\) determines the trace-vector part
\begin{equation}
\label{eq:torsion_trace_part}
{}^{\mathrm{Tr}}S^i{}_{jk}
:=
\frac13
\left(
\delta^i_kS_j-\delta^i_jS_k
\right).
\end{equation}
Next, with \(S_{ijk}:=g_{(i\ell)}S^\ell{}_{jk}\), the axial part is
obtained by complete antisymmetrization:
\begin{equation}
\label{eq:torsion_axial_part}
{}^{\mathrm A}S^i{}_{jk}
:=
\sginv^{i\ell}S_{[\ell jk]}
=
\frac16\varepsilon^i{}_{jkl}\widetilde S^l.
\end{equation}
This relation fixes the normalization of the axial vector
\(\widetilde S^l\) relative to the chosen Levi--Civita tensor.
The remaining part is defined by
\begin{equation}
\label{eq:torsion_pure_tensor_part}
{}^{0}S^i{}_{jk}
=
s^i{}_{jk}
:=
S^i{}_{jk}
-
{}^{\mathrm{Tr}}S^i{}_{jk}
-
{}^{\mathrm A}S^i{}_{jk}.
\end{equation}
It satisfies \(s^i{}_{ik}=0\) and \(s_{[ijk]}=0\), where
\(s_{ijk}:=g_{(i\ell)}s^\ell{}_{jk}\). Hence
\begin{equation}
\label{eq:torsion_decomposition}
S^i{}_{jk}
=
{}^{\mathrm{Tr}}S^i{}_{jk}
+
{}^{\mathrm A}S^i{}_{jk}
+
{}^{0}S^i{}_{jk},
\end{equation}
with \(24=4\oplus4\oplus16\) independent components.

\subsection{Pseudo-Hermitian transposition}

The simultaneous transposition of the metric and connection is the
component transformation used in the classical nonsymmetric theory
\cite{einstein_meaning_relativity,hlavaty1957}.

\begin{definition}[Pseudo-Hermitian invariance]
\label{def:pseudo_hermitian}
A Lagrangian density is pseudo-Hermitian invariant if its value is
unchanged under
\begin{equation}
\label{eq:ph}
g_{ij}\longmapsto g_{ji},\qquad
g^{ij}\longmapsto g^{ji},\qquad
\Gamma^i{}_{jk}\longmapsto\Gamma^i{}_{kj}.
\end{equation}
\end{definition}

In this paper, the term pseudo-Hermitian refers only to this discrete
transposition of the metric-affine variables. Its invariance for the
Lagrangian density introduced below will be verified explicitly.

\section{Einstein's Lagrangian Density and Variational Equations}
\label{sec:lagrangian}

We now formulate the variational system underlying Einstein's
nonsymmetric metric-affine theory. After introducing the relevant
curvature contractions and the Lagrangian density, we derive the
Euler--Lagrange equations by independent variation of the metric and
affine connection. We then examine the consequences of the connection
equation used in the subsequent analysis of the symmetric and
antisymmetric parts of the metric field equation.

\subsection{Curvature contractions and Lagrangian density}

We use the Lagrangian density of Einstein's late nonsymmetric
metric-affine theory in the notation introduced above; see
\cite{einstein_meaning_relativity,hlavaty1957}. The curvature convention
is
\begin{equation}
\label{eq:generalized_curvature_tensor}
R^n{}_{ikl}
:=
\Gamma^n{}_{il,k}
-
\Gamma^n{}_{ik,l}
+
\Gamma^n{}_{sk}\Gamma^s{}_{il}
-
\Gamma^n{}_{sl}\Gamma^s{}_{ik}.
\end{equation}
Its first and second contractions are
\begin{equation}
\label{eq:first_curvature_trace}
{}^{1}\!R_{ik}
:=
R^s{}_{isk}
=
\Gamma^s{}_{ik,s}
-
\Gamma^s{}_{is,k}
+
\Gamma^s{}_{ns}\Gamma^n{}_{ik}
-
\Gamma^s{}_{nk}\Gamma^n{}_{is},
\end{equation}
and
\begin{equation}
\label{eq:second_curvature_trace}
{}^{2}\!R_{kl}
:=
R^s{}_{skl}
=
\Gamma^s{}_{sl,k}
-
\Gamma^s{}_{sk,l}.
\end{equation}
Thus \({}^{2}\!R_{kl}\) is antisymmetric directly from its definition.

\begin{definition}[Lagrangian density]
\label{def:lagrangian_density}
We consider
\begin{equation}
\label{eq:L}
\mathcal L
=
\mathfrak g^{ji}P_{ij},
\end{equation}
where
\begin{equation}
\label{eq:Pij}
P_{ij}
:=
{}^{1}\!R_{ij}
-
A_{\substack{i\\+}\,\substack{;\,j\\\vphantom{+}}},
\end{equation}
and
\begin{equation}
\label{eq:Aplus}
A_{\substack{i\\+}}
:=
\frac{\omega_{,i}}{\omega}
-
\Gamma^s{}_{is}.
\end{equation}
\end{definition}

Equivalently,
\begin{equation}
\label{eq:Pik_explicit}
P_{ik}
=
\Gamma^s{}_{ik,s}
-
\Gamma^s{}_{nk}\Gamma^n{}_{is}
-
\left(
\frac{\omega_{,i}}{\omega}
\right)_{,k}
+
\frac{\omega_{,n}}{\omega}\Gamma^n{}_{ik}.
\end{equation}
The ordering of the lower indices in the last connection coefficient is
intentional. The second curvature trace does not enter
\(\mathcal L\) directly, but will appear through the generalized Bianchi
identities.

\begin{proposition}[Pseudo-Hermitian invariance]
\label{prop:pseudo_hermitian_L}
The Lagrangian density \eqref{eq:L} is invariant under the simultaneous
transposition \eqref{eq:ph}.
\end{proposition}

\begin{proof}
Under \eqref{eq:ph}, the determinant \(g\), and hence \(\omega\), is
unchanged, while
\(\mathfrak g^{ij}\mapsto\mathfrak g^{ji}\) and
\(P_{ij}\mapsto P_{ji}\); see
\cite{einstein_meaning_relativity,stolin1999_dissertation}. Therefore
\[
\mathcal L
=
\mathfrak g^{ji}P_{ij}
\longmapsto
\mathfrak g^{ij}P_{ji}
=
\mathfrak g^{ji}P_{ij}.
\]
\end{proof}

\subsection{Euler--Lagrange equations}

For the action
\begin{equation}
\label{eq:action}
\mathcal S[g,\nabla]
:=
\int_{\Omega}\mathcal L\,\dd^4x,
\end{equation}
the metric and the affine connection are varied independently, with
compactly supported variations. In particular, the metric variations
are not restricted by
$\partial_j\delta\mathfrak g^{[ij]}=0$; this is the precise sense of
``unrestricted'' used here.

\begin{proposition}[Euler--Lagrange equations]
\label{prop:EL_systems}
Independent variations of \(g\) and \(\nabla\) in \eqref{eq:action}
give the following two Euler--Lagrange equations. The metric variation
gives
\begin{equation}
\label{eq:EL1}
P_{vu}
-
\frac12(P-Q)g_{vu}
=
0,
\end{equation}
where
\begin{equation}
\label{eq:Q_scalar_definition}
P:=g^{uv}P_{vu},
\qquad
Q
:=
\frac{1}{\omega}
\mathcal Q^w{}_{\substack{;\,w\\0}},
\end{equation}
and
\begin{equation}
\label{eq:Qw}
\mathcal Q^w
:=
\mathfrak g^{\substack{(x\\-}\,\substack{w)\\+}}{}_{;\,x}
+
\mathfrak g^{[kw]}S_k.
\end{equation}
The connection variation gives
\begin{equation}
\label{eq:EL2}
\omega\,
g^{\substack{k\\-}\,\substack{i\\+}}{}_{;\,l}
=
0.
\end{equation}
\end{proposition}

\begin{proof}
The equations follow by unrestricted compactly supported variations of
the nonsymmetric metric and the affine connection, followed by
integration by parts. A complete component derivation is given in
\cite{stolin1999_dissertation}.
\end{proof}

\begin{corollary}[Equivalent form of the metric equation]
\label{cor:EL1_equivalent_form}
The trace and the equivalent form of \eqref{eq:EL1} are
\begin{equation}
\label{eq:P_trace_Q}
P=2Q,
\end{equation}
and
\begin{equation}
\label{eq:EL1_Pik_form}
P_{ik}
=
\frac12Q\,g_{ik},
\end{equation}
respectively.
\end{corollary}

\begin{proof}
Contracting \eqref{eq:EL1} with \(g^{uv}\) and using
\(g^{uv}g_{vu}=4\) gives \eqref{eq:P_trace_Q}; substitution back into
\eqref{eq:EL1} gives \eqref{eq:EL1_Pik_form}.
\end{proof}

\subsection{Consequences of the connection equation}

We collect the consequences of the connection equation
\eqref{eq:EL2} used below in the symmetric and antisymmetric parts of
the metric Euler--Lagrange equations.

Using the inverse-matrix identity, Eq.~\eqref{eq:EL2} is
equivalent to
\begin{equation}
\label{eq:EL2_covariant_form}
g_{\substack{k\\+}\,\substack{i\\-}\,
  \substack{;\,l\\\vphantom{+}}}
=
0,
\end{equation}
or, in components,
\begin{equation}
\label{eq:EL2_component_form}
g_{ki,l}
-
g_{si}\Gamma^s{}_{kl}
-
g_{ks}\Gamma^s{}_{li}
=
0.
\end{equation}

\begin{lemma}[Metric-density consequences]
\label{lem:metric_density_consequence_EL2}
If \eqref{eq:EL2} holds, then
\begin{align}
\omega_{\substack{;\,l\\0}}
&=0,
\label{eq:omega_covariant_derivative_zero}
\\
\mathfrak g^{\substack{(x\\-}\,\substack{w)\\+}}{}_{;\,x}
&=0,
\label{eq:symmetric_metric_density_derivative_zero}
\\
\mathfrak g^{[ti]}{}_{,t}
&=
\mathfrak g^{(ti)}\Gamma^s{}_{[ts]}
=
\mathfrak g^{(ti)}S_t.
\label{eq:antisym_density_relation_A}
\end{align}
\end{lemma}

\begin{proof}
Contracting \eqref{eq:EL2_covariant_form} with \(g^{ik}\) and using
\[
g^{ik}
g_{\substack{i\\+}\,\substack{k\\-}\,
  \substack{;\,l\\\vphantom{+}}}
=
2\frac{\omega_{\substack{;\,l\\0}}}{\omega}
\]
gives \eqref{eq:omega_covariant_derivative_zero}. The symmetric and
antisymmetric contractions of the corresponding density relation give
\eqref{eq:symmetric_metric_density_derivative_zero} and
\eqref{eq:antisym_density_relation_A}.
\end{proof}

\begin{corollary}[Reduction of the \(Q\)-term]
\label{cor:Q_reduction_EL2}
If \eqref{eq:EL2} holds, then
\begin{align}
\mathcal Q^w
&=
\mathfrak g^{[kw]}S_k,
\label{eq:Q_density_reduced_EL2}
\\
Q
&=
\left(
g^{[kw]}S_k
\right)_{\substack{;\,w\\0}}.
\label{eq:Q_scalar_reduced_EL2}
\end{align}
\end{corollary}

\begin{proof}
Equations~\eqref{eq:symmetric_metric_density_derivative_zero},
\eqref{eq:Qw}, and \eqref{eq:omega_covariant_derivative_zero}, together
with \(\mathfrak g^{[kw]}=\omega g^{[kw]}\), give
\eqref{eq:Q_density_reduced_EL2} and
\eqref{eq:Q_scalar_reduced_EL2}.
\end{proof}

For every contravariant vector density \(\mathcal V^i\) of weight one,
contraction of \eqref{eq:vector_density_derivative_zero} gives
\begin{equation}
\label{eq:vector_density_divergence}
\mathcal V^i{}_{\substack{;\,i\\0}}
=
\mathcal V^i{}_{,i},
\end{equation}
since the two connection terms cancel after relabelling dummy indices.

Define the torsion-trace density by
\begin{equation}
\label{eq:A_density_definition}
\mathcal A^i
:=
\mathfrak g^{(ti)}S_t.
\end{equation}
Equation~\eqref{eq:antisym_density_relation_A} gives
\(\mathcal A^i=\mathfrak g^{[ti]}{}_{,t}\). Hence antisymmetry and
\eqref{eq:vector_density_divergence} imply
\begin{equation}
\label{eq:A_density_conservation}
\mathcal A^i{}_{\substack{;\,i\\0}}
=
\mathcal A^i{}_{,i}
=
0.
\end{equation}
This is a Lorenz-type condition for the torsion-trace density; it
follows from the connection equation rather than being imposed
independently.

In Sections~\ref{sec:symmetric_part_first_EL_system} and
\ref{sec:antisymmetric_part_first_EL_system}, both field equations
\eqref{eq:EL1} and \eqref{eq:EL2} are understood unless a statement
explicitly requires only the connection equation. Additional reductions
are stated separately. The symmetric and antisymmetric equations are
parts of one variational system, not independent subsystems.

\section{Symmetric Part of the Metric Euler--Lagrange Equation}
\label{sec:symmetric_part_first_EL_system}

Using the connection equation \eqref{eq:EL2}, we decompose the
symmetric part of the metric Euler--Lagrange equation
\eqref{eq:EL1} into its Levi--Civita and non-Riemannian contributions.

\subsection{Curvature decomposition and generalized gravitational equations}

The connection equation \eqref{eq:EL2} admits the classical form used
in the nonsymmetric metric-affine framework of Hlavat{\'y}
\cite{hlavaty1957},
\begin{equation}
\label{eq:connection_decomposition_U}
\Gamma^i{}_{(jk)}
=
\lc{i}{jk}
+
U^i{}_{jk},
\end{equation}
where
\begin{equation}
\label{eq:Uijk_definition}
U^i{}_{jk}
:=
\sginv^{is}
\left(
\Gamma^t{}_{[sj]}g_{[kt]}
+
\Gamma^t{}_{[sk]}g_{[jt]}
\right).
\end{equation}
Thus \(U^i{}_{jk}=U^i{}_{kj}\). Related formulations appear also in
the work of Tonnelat \cite{tonnelat1966}.

On the connection field equation,
\eqref{eq:omega_covariant_derivative_zero} gives
\(A_{\substack{i\\+}}=-S_i\). Substitution into \eqref{eq:Pij} then
yields the decomposition of \(P_{ik}\)
\[
P_{ik}
=
\widehat R_{ik}
+
S^s{}_{ik\substack{;\,s\\0}}
+
\Sigma_{ik},
\qquad
\widehat R_{ik}
=
R_{ik}
+
U_{ik},
\]
where \(\widehat R_{ik}\) denotes the Ricci tensor of the symmetrized
affine connection with coefficients \(\Gamma^i{}_{(jk)}\). Hence
\[
U_{ik}
=
\widehat R_{ik}
-
R_{ik},
\]
so that \(U_{ik}\) is the tensorial difference between the Ricci tensor
of the symmetrized affine connection and that of the Levi--Civita
connection associated with \(g_{(ij)}\).

The tensor \(\widehat R_{ik}\) is symmetric because
\(\Gamma^s{}_{(si)}=\partial_i\ln\omega\), and \(\Sigma_{ik}\) is
symmetric by its quadratic torsion form below. The torsion divergence
is antisymmetric. The symmetric part therefore gives
\begin{equation}
\label{eq:P_sym_decomposition}
P_{(ik)}
=
R_{ik}
+
U_{ik}
+
\Sigma_{ik},
\end{equation}
where \(R_{ik}\) are the components of the Ricci tensor associated with
the Levi--Civita connection of \(g_{(ij)}\),
\begin{equation}
\label{eq:Rik_levicivita_part}
R_{ik}
=
{\lc{s}{ik}}_{,s}
-
{\lc{s}{is}}_{,k}
+
\lc{s}{ns}\lc{n}{ik}
-
\lc{s}{nk}\lc{n}{is},
\end{equation}
while the Ricci-difference tensor \(U_{ik}\) is
\begin{equation}
\label{eq:Uik_connection_tensor_part}
U_{ik}
=
U^s{}_{ik;\,s}
-
U^s{}_{is;\,k}
+
U^s{}_{ns}U^n{}_{ik}
-
U^s{}_{nk}U^n{}_{is},
\end{equation}
where both semicolons denote Levi--Civita covariant derivatives, and
\begin{equation}
\label{eq:Sigmaik_torsion_term}
\Sigma_{ik}
=
-\Gamma^n{}_{[is]}\Gamma^s{}_{[nk]}.
\end{equation}
We denote
\begin{equation}
\label{eq:RUSigma_traces}
R\equiv \sginv^{ij}R_{ij},
\qquad
U\equiv \sginv^{ij}U_{ij},
\qquad
\Sigma\equiv \sginv^{ij}\Sigma_{ij}.
\end{equation}

\begin{proposition}[Generalized gravitational equations]
\label{prop:generalized_grav_equations}
The symmetric metric equation can be written in the Einstein-type form
\begin{equation}
\label{eq:generalized_grav_equation}
R_{ij}
-
\tfrac12 R\,g_{(ij)}
=
T_{(ij)},
\end{equation}
where
\begin{equation}
\label{eq:geometric_grav_source_definition}
T_{(ij)}
:=
-
\left(
U_{ij}
-
\tfrac12 U\,g_{(ij)}
+
\Sigma_{ij}
-
\tfrac12 \Sigma\,g_{(ij)}
+
\tfrac12 Q\,g_{(ij)}
\right).
\end{equation}
The geometrically induced source term satisfies
\begin{equation}
\label{eq:geometric_grav_source_conservation}
T^{(ij)}{}_{;\,j}=0,
\end{equation}
where indices are raised with \(g_{(ij)}\).
\end{proposition}

\begin{proof}
Taking the symmetric part of \eqref{eq:EL1} and using
\eqref{eq:P_sym_decomposition} gives
\eqref{eq:generalized_grav_equation}, with \(T_{(ij)}\) defined by
\eqref{eq:geometric_grav_source_definition}.

The contracted Bianchi identity for
\[
G_{ij}:=R_{ij}-\tfrac12R\,g_{(ij)}
\]
gives \(G^{ij}{}_{;\,j}=0\); see, for example,
Landau and Lifshitz~\cite{landau_lifshitz_fields}.
Equation~\eqref{eq:geometric_grav_source_conservation} then follows
from \eqref{eq:generalized_grav_equation}.
\end{proof}

\subsection{Vacuum Einstein limit}

Consider the torsion-free reduction
\begin{equation}
\label{eq:torsion_free_reduction}
\Gamma^i{}_{[jk]}=0.
\end{equation}
Then \(S^i{}_{jk}=0\), and
\eqref{eq:Uijk_definition} gives \(U^i{}_{jk}=0\). Hence
\eqref{eq:connection_decomposition_U}, together with
\eqref{eq:torsion_free_reduction}, yields
\begin{equation}
\label{eq:levi_civita_reduction}
\Gamma^i{}_{jk}
=
\lc{i}{jk}.
\end{equation}

\begin{corollary}[Vacuum Einstein equation in the torsion-free reduction]
\label{cor:vacuum_einstein_limit}
Under the torsion-free reduction \eqref{eq:torsion_free_reduction},
\begin{equation}
\label{eq:vacuum_einstein_equations}
R_{ij}=0.
\end{equation}
\end{corollary}

\begin{proof}
Equation~\eqref{eq:levi_civita_reduction} gives \(U_{ij}=0\), while
\eqref{eq:torsion_free_reduction} gives \(\Sigma_{ij}=0\) and
\(S_i=0\). Corollary~\ref{cor:Q_reduction_EL2} therefore gives
\(Q=0\), so that \(T_{(ij)}=0\). Equation
\eqref{eq:generalized_grav_equation} reduces to
\[
R_{ij}-\tfrac12R\,g_{(ij)}=0.
\]
Taking the trace in four dimensions gives \(R=0\), and hence
\eqref{eq:vacuum_einstein_equations}.
\end{proof}

The antisymmetric part \(g_{[ij]}\) is not constrained by the reduced
symmetric gravitational equation.

\section{Antisymmetric Part of the Metric Euler--Lagrange Equation}
\label{sec:antisymmetric_part_first_EL_system}

We now analyze the antisymmetric part of the metric Euler--Lagrange
equation \eqref{eq:EL1}, together with the connection equation
\eqref{eq:EL2}. The resulting structure consists of a homogeneous
Maxwell-type identity associated with the torsion trace, a geometric
constitutive relation, and its contracted four-divergence. The
Maxwell-type and Proca-type equations arise as particular reductions of
the latter.

\subsection{Trace of the generalized Bianchi identities}

We use the generalized curvature tensor
\eqref{eq:generalized_curvature_tensor} and the second curvature trace
\eqref{eq:second_curvature_trace}. Generalized Bianchi identities play
a central role already in Einstein's nonsymmetric theory and in recent
analyses of its differential identities
\cite{einstein1950_bianchi,wanas_nabil_abdelhamid_elabd2025}; a component
derivation in closely related notation is given in
\cite[Theorem~1.5, p.~22]{stolin1999_dissertation}.

Let \(\mathcal P:=\{+,-,0\}\), and let
\(\sigma_i\in\mathcal P\) for \(i=1,2,3\). Define
\(\sigma_i^*\in\mathcal P\) by
\begin{equation}
\label{eq:sigma_star_definition}
\sigma_i^*
=
\begin{cases}
+,& \text{if } \sigma_i=-,\\
-,& \text{if } \sigma_i=+,\\
0,& \text{if } \sigma_i=0,
\end{cases}
\qquad i=1,2,3.
\end{equation}
The generalized Bianchi identity is written as
\begin{equation}
\label{eq:generalized_bianchi_identity}
R^{\substack{n\\+}}{}_{\substack{i\\+}\,\substack{k\\\sigma_1}\,
\substack{l\\\sigma_2};\,m}
+
R^{\substack{n\\+}}{}_{\substack{i\\+}\,\substack{m\\\sigma_2^*}\,
\substack{k\\\sigma_3^*};\,l}
+
R^{\substack{n\\+}}{}_{\substack{i\\+}\,\substack{l\\\sigma_3}\,
\substack{m\\\sigma_1^*};\,k}
=
0 .
\end{equation}

\begin{proposition}[Trace of the generalized Bianchi identities]
\label{prop:maxwell_bianchi}
If the connection equation \eqref{eq:EL2} holds, then the components of
the second curvature trace satisfy
\begin{equation}
\label{eq:second_curvature_trace_torsion}
{}^{2}\!R_{ij}
=
S_{i,j}-S_{j,i}
=
-2\partial_{[i}S_{j]}.
\end{equation}
Consequently,
\begin{equation}
\label{eq:maxwell_bianchi}
{}^{2}\!R_{ij,k}
+
{}^{2}\!R_{jk,i}
+
{}^{2}\!R_{ki,j}
=
0.
\end{equation}
\end{proposition}

\begin{proof}
By Lemma~\ref{lem:metric_density_consequence_EL2},
\eqref{eq:EL2} implies
\(\Gamma^s{}_{(si)}=(\ln\omega)_{,i}\). Since
\(S_i=\Gamma^s{}_{[is]}\),
\[
\Gamma^s{}_{si}
=
(\ln\omega)_{,i}-S_i.
\]
Substitution into \eqref{eq:second_curvature_trace} gives
\eqref{eq:second_curvature_trace_torsion}. Contracting the generalized
Bianchi identity \eqref{eq:generalized_bianchi_identity} then yields
\eqref{eq:maxwell_bianchi}.
\end{proof}

\begin{remark}[Electromagnetic interpretation]
\label{rem:em_interpretation_torsion_trace}
Equation~\eqref{eq:second_curvature_trace_torsion} has the same formal
structure as a field strength generated by a vector potential. Thus the
torsion trace \(S_i\) has a potential-type role and
\({}^{2}\!R_{ij}\) a field-strength-type role, while
\eqref{eq:maxwell_bianchi} is the corresponding homogeneous
Maxwell-type identity. This interpretation is also compatible with the
Lorenz-type condition \eqref{eq:A_density_conservation}. No independent
electromagnetic field is introduced.
\end{remark}

\subsection{Torsion-divergence decomposition and constitutive equation}

Taking the antisymmetric part of \(P_{ik}\) and using
\eqref{eq:Aplus}, together with
\eqref{eq:omega_covariant_derivative_zero}, gives
\begin{equation}
\label{eq:P_antisym_connection_form}
P_{[ik]}
=
\Gamma^j{}_{[ik]\substack{;\,j\\0}} .
\end{equation}

\begin{lemma}[Decomposition of the torsion divergence]
\label{lem:torsion_divergence_decomposition}
With \(S^i{}_{jk}=\Gamma^i{}_{[jk]}\) and
\(\Gamma^i{}_{(jk)}=\lc{i}{jk}+U^i{}_{jk}\), one has
\begin{equation}
\label{eq:torsion_divergence_decomposition_basic}
\Gamma^s{}_{[ik]\substack{;\,s\\0}}
=
S^s{}_{ik;\,s}
-
\left(S^s{}_{ik}\right)_{\substack{;\,s\\\Delta}}.
\end{equation}
Moreover,
\begin{equation}
\label{eq:ordinary_part_decomposition}
S^s{}_{ik;\,s}
=
\tfrac13
\left(
S_{i;\,k}
-
S_{k;\,i}
\right)
+
{}^{\mathrm A}S^s{}_{ik;\,s}
+
{}^{0}S^s{}_{ik;\,s},
\end{equation}
and
\begin{align}
\label{eq:Delta_part_decomposition}
\left(S^s{}_{ik}\right)_{\substack{;\,s\\\Delta}}
&=
S^s{}_{tk}U^t{}_{is}
+
S^s{}_{it}U^t{}_{ks}
-
S^t{}_{ik}U^s{}_{ts}
\nonumber\\
&=
\left({}^{\mathrm A}S^s{}_{tk}+{}^{0}S^s{}_{tk}\right)U^t{}_{is}
+
\left({}^{\mathrm A}S^s{}_{it}+{}^{0}S^s{}_{it}\right)U^t{}_{ks}
-
\left({}^{\mathrm A}S^t{}_{ik}+{}^{0}S^t{}_{ik}\right)U^s{}_{ts}.
\end{align}
\end{lemma}

\begin{proof}
Equation~\eqref{eq:torsion_divergence_decomposition_basic} follows from
\eqref{eq:Delta_derivative_definition}. For
\eqref{eq:ordinary_part_decomposition}, insert the irreducible
decomposition \eqref{eq:torsion_decomposition} into
\(S^s{}_{ik;\,s}\) and use \eqref{eq:torsion_trace_part}.
Equation~\eqref{eq:Delta_part_decomposition} follows by applying the
difference between the Levi--Civita and symmetrized affine connections
to all three indices of \(S^s{}_{ik}\).
\end{proof}

The decomposition separates the trace contribution
\(\tfrac13(S_{i;\,k}-S_{k;\,i})\) from the
\(\Delta\)-correction. By \eqref{eq:Delta_part_decomposition}, the
latter contains no explicit trace-torsion factor and is bilinear in
\(U^i{}_{jk}\) and the axial and traceless torsion components.
Together with
\[
U_{ij}=\widehat R_{ij}-R_{ij},
\]
this exhibits the same connection-difference principle in the
symmetric curvature and antisymmetric divergence terms: in the former,
\(U_{ij}\) measures the difference between the Ricci tensors of the
symmetrized affine and Levi--Civita connections, whereas in the latter
the \(\Delta\)-term measures the corresponding difference between their
covariant divergences of torsion. The splitting is additive, but its
corrections contain products of geometric fields and are not independent
sources. In particular,
\[
U^i{}_{jk}=0
\quad\Longrightarrow\quad
\left(S^s{}_{ik}\right)_{\substack{;\,s\\\Delta}}=0.
\]

\begin{definition}[Antisymmetric constitutive tensors]
\label{def:antisymmetric_constitutive_tensors}
We define antisymmetric tensor fields by their components
\begin{equation}
\label{eq:H_tensor_definition}
H_{ik}
:=
S^s{}_{ik;\,s},
\end{equation}
\begin{equation}
\label{eq:F_tensor_trace_definition}
F_{ik}
:=
\frac13
\left(
S_{i;\,k}
-
S_{k;\,i}
\right),
\end{equation}
\begin{equation}
\label{eq:D_axial_tensor_definition}
{}^{\mathrm A}D_{ik}
:=
-{}^{\mathrm A}S^s{}_{ik;\,s},
\end{equation}
\begin{equation}
\label{eq:D_pure_tensor_definition}
{}^{0}D_{ik}
:=
-{}^{0}S^s{}_{ik;\,s},
\end{equation}
and
\begin{equation}
\label{eq:J_tensor_definition}
J_{ik}
:=
\left(S^s{}_{ik}\right)_{\substack{;\,s\\\Delta}}.
\end{equation}
We further define
\begin{equation}
\label{eq:D_tensor_definition}
D_{ik}
:=
{}^{\mathrm A}D_{ik}
+
{}^{0}D_{ik}.
\end{equation}
\end{definition}
By \eqref{eq:F_tensor_trace_definition} and
\eqref{eq:second_curvature_trace_torsion}, and using the symmetry of
the Levi--Civita connection,
\begin{equation}
\label{eq:F_second_curvature_relation}
F_{ik}
=
\frac13\,{}^{2}\!R_{ik}.
\end{equation}
Thus the field-strength-type contribution entering the constitutive
relation is directly proportional to the same curvature trace whose
cyclic identity gives the homogeneous Maxwell-type equation
\eqref{eq:maxwell_bianchi}.

All these tensor fields are antisymmetric. Equations
\eqref{eq:Delta_part_decomposition} and \eqref{eq:J_tensor_definition}
also show that \(J_{ik}\) contains no explicit trace-torsion factor.
In particular,
\[
{}^{\mathrm A}S^i{}_{jk}={}^{0}S^i{}_{jk}=0
\quad\Longrightarrow\quad
J_{ik}=0,
\]
and \(U^i{}_{jk}=0\) likewise implies \(J_{ik}=0\).

\begin{remark}[Constitutive interpretation]
\label{rem:constitutive_interpretation}
The components \(F_{ik}\), \(D_{ik}\), and \(H_{ik}\) have, respectively, field-strength-type, polarization--magnetization-type response, and constitutive-resultant roles in the interpretation used below. These identifications are structural and do not imply that the corresponding quantities coincide with independently introduced electromagnetic field, polarization, or excitation tensors of standard electrodynamics. The tensor \(J_{ik}\) collects precisely the contribution associated with the difference between the symmetrized affine and Levi--Civita connections and gives rise to a source-like contribution after taking the contracted four-divergence. All these quantities are composite geometric expressions obtained from the underlying metric-affine variables.
\end{remark}

\begin{proposition}[Antisymmetric metric Euler--Lagrange equation]
\label{prop:antisym_EL_part}
The antisymmetric metric equation has the constitutive form
\begin{equation}
\label{eq:H_constitutive_tensor_relation}
H_{ik}
=
F_{ik}
-
D_{ik}
=
T_{[ik]},
\end{equation}
where
\begin{equation}
\label{eq:antisymmetric_source_tensor_definition}
T_{[ik]}
:=
J_{ik}
+
\frac12 Q\,g_{[ik]} .
\end{equation}
\end{proposition}

\begin{proof}
The antisymmetric part of \eqref{eq:EL1_Pik_form} gives
\[
P_{[ik]}=\frac12Q\,g_{[ik]}.
\]
Using \eqref{eq:P_antisym_connection_form} and
\eqref{eq:torsion_divergence_decomposition_basic},
\[
P_{[ik]}
=
H_{ik}-J_{ik}.
\]
Hence
\[
H_{ik}
=
J_{ik}
+
\frac12Q\,g_{[ik]}.
\]
On the other hand, \eqref{eq:ordinary_part_decomposition} and
Definition~\ref{def:antisymmetric_constitutive_tensors} give
\(H_{ik}=F_{ik}-D_{ik}\), which proves
\eqref{eq:H_constitutive_tensor_relation}.
\end{proof}

\subsection{Four-divergence of the constitutive relation}

The contracted four-divergence has a structural role in the present
construction. The antisymmetric metric Euler--Lagrange equation first
takes a constitutive form, and its Levi--Civita divergence therefore
passes from the constitutive level to the corresponding vector-field
level. A second divergence gives the associated conservation identity.
This is the antisymmetric counterpart of the Einstein-type equation and
its covariant conservation law in the symmetric part, but shifted by
one differential level.

All indices in this subsection are raised with the symmetric metric.
For any antisymmetric contravariant tensor \(X^{ik}\),
Eq.~\eqref{eq:LC_divergence_volume_form} gives its Levi--Civita
divergence. The same expression written with the full volume density is
related to it, on the connection equation, by
\begin{equation}
\label{eq:full_vs_symmetric_volume_divergence}
\frac{1}{\omega}\partial_k\!\left(\omega X^{ik}\right)
=
X^{ik}{}_{;\,k}+U_kX^{ik},
\qquad
U_k:=U^s{}_{sk},
\end{equation}
because
\(\partial_k\ln(\omega/\omega_{\mathrm S})=U_k\).
Equation~\eqref{eq:full_vs_symmetric_volume_divergence} is the only
conversion between the two volume elements needed below.

\begin{lemma}[Vanishing divergence of the axial contribution]
\label{lem:D_axial_divergence_zero}
For the axial tensor defined in
\eqref{eq:D_axial_tensor_definition},
\begin{equation}
\label{eq:D_axial_divergence_zero}
{}^{\mathrm A}D^{ik}{}_{;\,k}=0.
\end{equation}
\end{lemma}

\begin{proof}
Raise all indices with the symmetric metric. By Levi--Civita metric
compatibility and total antisymmetry of \({}^{\mathrm A}S^{sik}\),
\[
\omega_{\mathrm S}\,{}^{\mathrm A}D^{ik}
=
-\partial_s\!\left(\omega_{\mathrm S}
{}^{\mathrm A}S^{sik}\right).
\]
Using \eqref{eq:LC_divergence_volume_form} once more,
\[
{}^{\mathrm A}D^{ik}{}_{;\,k}
=
-\frac{1}{\omega_{\mathrm S}}
\partial_k\partial_s\!\left(\omega_{\mathrm S}
{}^{\mathrm A}S^{sik}\right)
=0,
\]
by antisymmetry in \(s,k\) and commutativity of partial derivatives.
\end{proof}

For later use define only the raised antisymmetric metric tensor
\begin{equation}
\label{eq:raised_antisymmetric_metric_tensor_definition}
B^{ik}
:=
g_{[mn]}\,{}^{\mathrm S}g^{im}\,{}^{\mathrm S}g^{kn}.
\end{equation}
Taking the antisymmetric part of
\eqref{eq:EL2_covariant_form} and contracting with
\(\sginv^{kl}\) gives
\begin{equation}
\label{eq:g_antisym_trace_connection_vector}
g_{[ik]\substack{;\,l\\0}}
\,\sginv^{kl}
=
S_i.
\end{equation}
Comparison of the Levi--Civita and symmetrized affine derivatives then
gives
\begin{equation}
\label{eq:B_tensor_divergence_decomposition}
B^{ik}{}_{;\,k}
=
S^i+C^i,
\end{equation}
where
\begin{equation}
\label{eq:C_tensor_correction_definition}
C^i
:=
\left(
U^s{}_{mk}g_{[sn]}
+U^s{}_{nk}g_{[ms]}
\right)
\sginv^{im}\sginv^{kn}.
\end{equation}
The correction \(C^i\) is linear in the connection difference
\(U^i{}_{jk}\) and vanishes when that difference vanishes.

\begin{proposition}[Levi--Civita four-divergence of the antisymmetric constitutive relation]
\label{prop:four_divergence_antisym_part}
After raising indices with the symmetric metric, the contracted
Levi--Civita divergence of the constitutive relation
\eqref{eq:H_constitutive_tensor_relation} gives
\begin{equation}
\label{eq:general_Maxwell_Proca_divergence}
H^{ik}{}_{;\,k}
=
F^{ik}{}_{;\,k}
-
{}^{0}D^{ik}{}_{;\,k}
=
T^{[ik]}{}_{;\,k},
\end{equation}
where
\begin{equation}
\label{eq:antisymmetric_source_divergence_expanded}
\begin{aligned}
T^{[ik]}{}_{;\,k}
&=
J^{ik}{}_{;\,k}
+\tfrac12 Q_{,k}B^{ik}
+\tfrac12Q\,B^{ik}{}_{;\,k}
\\
&=
J^{ik}{}_{;\,k}
+\tfrac12 Q_{,k}B^{ik}
+\tfrac12Q\left(S^i+C^i\right).
\end{aligned}
\end{equation}
Moreover, since \(T^{[ik]}\) is antisymmetric, a further contracted
Levi--Civita divergence vanishes identically:
\begin{equation}
\label{eq:antisymmetric_current_conservation}
\left(T^{[ik]}{}_{;\,k}\right)_{;\,i}=0.
\end{equation}
\end{proposition}

\begin{proof}
Raise the tensor relation
\[
H_{ik}=F_{ik}-{}^{\mathrm A}D_{ik}-{}^{0}D_{ik}
=T_{[ik]}
\]
with the symmetric metric and take its Levi--Civita divergence.
Lemma~\ref{lem:D_axial_divergence_zero} gives
\eqref{eq:general_Maxwell_Proca_divergence}. Equations
\eqref{eq:antisymmetric_source_tensor_definition} and
\eqref{eq:B_tensor_divergence_decomposition}, together with the product
rule, give \eqref{eq:antisymmetric_source_divergence_expanded}.
Finally, by \eqref{eq:LC_divergence_volume_form},
\[
\left(T^{[ik]}{}_{;\,k}\right)_{;\,i}
=
\frac{1}{\omega_{\mathrm S}}
\partial_i\partial_k
\left(\omega_{\mathrm S}T^{[ik]}\right)
=
0,
\]
by antisymmetry in \(i,k\) and commutativity of partial derivatives.
\end{proof}

Thus the axial contribution remains in the constitutive relation but
drops out of its Levi--Civita divergence. The traceless tensor
contribution generally remains. If the same antisymmetric tensor is
instead written with the full volume element, the additional
\(U_kX^{ik}\) term is restored exactly by
\eqref{eq:full_vs_symmetric_volume_divergence}; no geometric content is
lost by using the tensor form above.

\subsection{Maxwell and Proca-type vacuum limits}

Here Maxwell-type and Proca-type refer to the forms of the reduced
vector equations, not to an established particle spectrum. The term
vacuum denotes their source-free form and should not be confused with
the torsion-free vacuum Einstein limit. In particular, the torsion trace and
the axial part of torsion may remain nonzero.

We consider the trace--axial reduction
\begin{equation}
\label{eq:em_vacuum_trace_axial_reduction}
U^i{}_{jk}=0,
\qquad
{}^{0}S^i{}_{jk}=0.
\end{equation}
The first condition is equivalent, through
\eqref{eq:connection_decomposition_U}, to
\[
\Gamma^i{}_{(jk)}=\lc{i}{jk},
\]
and therefore removes the \(\Delta\)-correction \(J_{ik}\). The second
condition removes the traceless tensor part of torsion, while its trace
and axial parts may remain.

The torsion-free condition \eqref{eq:torsion_free_reduction} is a
special case of \eqref{eq:em_vacuum_trace_axial_reduction}, since then
all irreducible torsion components vanish and
\eqref{eq:Uijk_definition} gives \(U^i{}_{jk}=0\).

The reduction \eqref{eq:em_vacuum_trace_axial_reduction} nevertheless
does not force the torsion tensor itself to vanish. Using
\eqref{eq:Uijk_definition} and
\[
S^i{}_{jk}
=
{}^{\mathrm{Tr}}S^i{}_{jk}
+
{}^{\mathrm A}S^i{}_{jk},
\]
one obtains the algebraic constraint
\begin{equation}
\label{eq:trace_axial_reduction_constraint}
\frac13
\left(
S_jg_{[rk]}
+
S_kg_{[rj]}
\right)
+
{}^{\mathrm A}S^t{}_{rj}g_{[kt]}
+
{}^{\mathrm A}S^t{}_{rk}g_{[jt]}
=
0.
\end{equation}
Thus the trace and axial torsion contributions are not independent in
the trace--axial reduction \eqref{eq:em_vacuum_trace_axial_reduction}.

In this reduction \(U_i=0\), and hence \(C^i=0\). Equation
\eqref{eq:B_tensor_divergence_decomposition} therefore becomes
\begin{equation}
\label{eq:B_tensor_divergence_trace_axial}
B^{ik}{}_{;\,k}=S^i.
\end{equation}
A further divergence gives the corresponding Levi--Civita Lorenz-type
condition
\begin{equation}
\label{eq:exact_lorenz_condition_trace_axial}
S^i{}_{;\,i}=0.
\end{equation}
This consequence should be distinguished from the general density
condition \eqref{eq:A_density_conservation}, which follows directly
from the connection equation before imposing the trace--axial
reduction. The homogeneous identity \eqref{eq:maxwell_bianchi} is
unaffected by the reduction.

\begin{proposition}[Maxwell- and Proca-type vacuum limits]
\label{prop:maxwell_proca_vacuum_limits_antisym_part}
Under the trace--axial reduction
\eqref{eq:em_vacuum_trace_axial_reduction}, the raised antisymmetric
constitutive relation reduces to
\begin{equation}
\label{eq:em_vacuum_constitutive_relation}
H^{ik}
=
F^{ik}
-{}^{\mathrm A}D^{ik}
=
\frac12Q B^{ik},
\end{equation}
and its contracted Levi--Civita divergence is
\begin{equation}
\label{eq:em_vacuum_maxwell_proca_general}
F^{ik}{}_{;\,k}
=
\frac12Q_{,k}B^{ik}
+
\frac12Q S^i.
\end{equation}

In particular:
\begin{enumerate}
\item If \(Q=0\), then
\begin{equation}
\label{eq:em_vacuum_maxwell_constitutive_relation}
H^{ik}
=
F^{ik}
-{}^{\mathrm A}D^{ik}
=
0,
\end{equation}
and
\begin{equation}
\label{eq:em_vacuum_maxwell_equation}
F^{ik}{}_{;\,k}=0.
\end{equation}

\item If \(Q=\mathrm{const}\neq0\), then
\begin{equation}
\label{eq:em_vacuum_proca_constitutive_relation}
H^{ik}
=
F^{ik}
-{}^{\mathrm A}D^{ik}
=
\frac12Q B^{ik},
\end{equation}
and
\begin{equation}
\label{eq:em_vacuum_proca_equation}
F^{ik}{}_{;\,k}
-
\frac12Q S^i
=
0.
\end{equation}
\end{enumerate}

In the case \(Q=0\), the right-hand side of
\eqref{eq:general_Maxwell_Proca_divergence} vanishes. For
\(Q=\mathrm{const}\neq0\), the remaining term \(\tfrac12QS^i\) has
the Proca-type potential role and is moved to the left-hand side of
\eqref{eq:em_vacuum_proca_equation}.
\end{proposition}

\begin{proof}
Under \eqref{eq:em_vacuum_trace_axial_reduction},
\(D_{ik}={}^{\mathrm A}D_{ik}\), \(J_{ik}=0\), \(U_i=0\), and
\(C^i=0\). Hence \eqref{eq:H_constitutive_tensor_relation} gives
\eqref{eq:em_vacuum_constitutive_relation}. Taking its Levi--Civita
divergence and using Lemma~\ref{lem:D_axial_divergence_zero} together
with \eqref{eq:B_tensor_divergence_trace_axial} gives
\eqref{eq:em_vacuum_maxwell_proca_general}.

For \(Q=0\), these equations give
\eqref{eq:em_vacuum_maxwell_constitutive_relation} and
\eqref{eq:em_vacuum_maxwell_equation}. For
\(Q=\mathrm{const}\neq0\), one has \(Q_{,k}=0\), which gives
\eqref{eq:em_vacuum_proca_constitutive_relation} and
\eqref{eq:em_vacuum_proca_equation}.
\end{proof}

The two cases above are sufficient reductions, not an exhaustive
classification. A nonzero source contribution may have vanishing
contracted four-divergence and therefore lead to the same divergence
equation without being equivalent to the \(Q=0\) case in the complete
Euler--Lagrange system.

\begin{remark}[Gauge interpretation of the Maxwell-type structure]
\label{rem:gauge_interpretation_maxwell_structure}
By \eqref{eq:second_curvature_trace_torsion}, the transformation
\[
S_i\longmapsto S_i+\partial_i\chi
\]
leaves the field-strength-type quantity \({}^{2}\!R_{ij}\) unchanged.
Thus the homogeneous equation \eqref{eq:maxwell_bianchi} has the usual
gauge-invariant character.

The condition \(Q=0\) removes the Proca-type contribution from
\eqref{eq:antisymmetric_source_tensor_definition}, but does not by
itself require \(J_{ij}=0\). If \(U^i{}_{jk}=0\), however,
\eqref{eq:J_tensor_definition} gives \(J_{ij}=0\). Hence for
\(Q=0\) and \(U^i{}_{jk}=0\),
\[
T_{[ij]}=0,
\qquad
H_{ij}=0,
\qquad
H^{ij}{}_{;\,j}=0.
\]
The axial and traceless torsion components may still enter the
constitutive structure when they are not removed by an additional
reduction.

This gauge interpretation concerns the reduced field-strength
structure and does not establish an off-shell gauge symmetry of the
complete metric-affine Euler--Lagrange system.
\end{remark}

\begin{corollary}[Torsion-free reduction of the antisymmetric equation]
\label{cor:antisymmetric_part_torsion_free_trivial}
Under the torsion-free reduction \eqref{eq:torsion_free_reduction},
the antisymmetric constitutive relation reduces to \(0=0\), its contracted
four-divergence is trivial, and the antisymmetric metric components
\(g_{[ij]}\) are not constrained by this reduced equation.
\end{corollary}

\begin{proof}
The torsion-free condition gives
\[
S^i{}_{jk}=0,
\qquad
H_{ik}=F_{ik}=D_{ik}=J_{ik}=0.
\]
Since \(S_i=0\), Corollary~\ref{cor:Q_reduction_EL2} gives \(Q=0\).
Thus \(T_{[ik]}=0\), and
\eqref{eq:H_constitutive_tensor_relation} is identically satisfied.
\end{proof}

The torsion-free Einstein limit is therefore distinct from the
Maxwell--Proca-type vacuum limits above: in the latter, the torsion trace
and axial torsion may remain nonzero, and the geometrically induced
gravitational source \(T_{(ij)}\) need not vanish.

\begin{remark}[Axial vacuum response]
\label{rem:axial_vacuum_response}
In the Maxwell-type vacuum limit,
\eqref{eq:em_vacuum_maxwell_constitutive_relation} permits a nonzero
axial contribution \({}^{\mathrm A}D^{ik}\), although its contracted
Levi--Civita divergence vanishes by
\eqref{eq:D_axial_divergence_zero}. If the axial contribution vanished
as well, the same constitutive relation would imply
\( F^{ik}=0\). Thus a nontrivial source-free Maxwell-type field
within this reduction requires a nonzero axial constitutive response.
This requirement is stronger than the divergence equation
\( F^{ik}{}_{;k}=0\) alone: the axial response is required for a
nonzero Maxwell-type configuration to remain compatible with the
underlying antisymmetric Euler--Lagrange constitutive relation.
\end{remark}

\section{Discussion and Conclusion}
\label{sec:discussion}

We have analyzed Einstein's nonsymmetric metric-affine Lagrangian
density with the metric and affine connection treated as independent
variational variables, without imposing the divergence constraint on the
antisymmetric metric density discussed in the Introduction. The connection
Euler--Lagrange equation belongs to the classical
Einstein--Hlavat{\'y}--Tonnelat framework, whereas the unrestricted metric
Euler--Lagrange equation provides both gravitational and
vector-field-type structures within the same variational system.

The symmetric part of the metric Euler--Lagrange equation can be written
as an Einstein-type gravitational equation for \(g_{(ij)}\), with a
geometrically induced source constructed from the non-Riemannian
components of the metric-affine geometry. In the torsion-free reduction,
the affine connection becomes the Levi--Civita connection, the geometric
source vanishes, and the vacuum Einstein equations are recovered. The
antisymmetric metric components are not constrained by this reduced
symmetric equation.

The antisymmetric part has a different but complementary structure.
Using the decomposition of torsion into irreducible components with
respect to the proper orthochronous component of the local Lorentz group
associated with \(g_{(ij)}\), it takes the form of a geometric constitutive relation.
Its contracted Levi--Civita divergence separates the contribution
associated with the torsion trace from the traceless torsion
contribution and from geometrically induced source-like terms. The
axial torsion contribution remains in the constitutive relation, while
its Levi--Civita divergence vanishes identically. The two natural
volume elements enter only through the exact conversion
\eqref{eq:full_vs_symmetric_volume_divergence}; they do not require two
parallel families of constitutive variables.

There is also a structural difference between the symmetric and
antisymmetric parts of the metric Euler--Lagrange equation. In the
symmetric part, the Einstein-type field equation is followed by the
covariant conservation law for its geometrically induced source. In the
antisymmetric part, the corresponding hierarchy is shifted by one
differential level: the metric Euler--Lagrange equation first has the
form of a constitutive relation. Its contracted four-divergence yields
the corresponding vector-field equation, whose further contracted
divergence vanishes identically and thus gives the associated
conservation identity. The last step follows directly from the
antisymmetry of the relevant rank-two tensor and has no analogous
automatic counterpart in the symmetric part. The divergence equation is
therefore a differential consequence of the antisymmetric metric
equation, not an equivalent replacement for it.

This organization differs from the usual macroscopic formulation of electrodynamics, in which Maxwell equations and constitutive relations are introduced as complementary ingredients and required to be mutually compatible. The distinction here is not the presence of a
constitutive structure itself, but its variational and differential
status: the antisymmetric metric Euler--Lagrange equation provides the
constitutive level, while the Maxwell--Proca-type dynamical level arises
as its contracted four-divergence.

Under the trace--axial reduction, the case \(Q=0\) gives a source-free
Maxwell-type equation, whereas \(Q=\mathrm{const}\neq0\) gives a
Proca-type equation. These are particular reductions of the general
antisymmetric equation and should not be regarded as an exhaustive
classification of all configurations having the same contracted
four-divergence.

The constitutive decomposition also distinguishes two qualitatively
different geometric response contributions. The traceless tensor part
of torsion gives a polarization--magnetization-like response, whereas
the axial contribution may remain nonzero even in the Maxwell-type
vacuum reduction. In fact, within that reduction a nontrivial
source-free field compatible with the full constitutive equation
requires a nonvanishing axial response. The Maxwell-type vacuum is
therefore not a structureless absence of constitutive response, but may
carry an intrinsic geometric constitutive structure.

This vacuum response has an interesting but only qualitative parallel
in quantum-field-theoretic studies of vacuum spin currents and
spin-polarization effects induced by electromagnetic backgrounds
\cite{chu_leung2021_spin_vacuum,
huang_matsuo_taya2019_spin_current_vacuum}. The mechanisms are different
and no direct equivalence is implied: the present result is a
constitutive compatibility condition and does not by itself establish a
dynamical mechanism for electromagnetic-wave propagation.

Independently of this constitutive reduction, the second curvature trace
and the generalized Bianchi identities give the homogeneous Maxwell-type
identity associated with the torsion trace, while the connection
Euler--Lagrange equation implies a Lorenz-type condition for the
corresponding torsion-trace density. Taken together, the results exhibit
a symmetry-organized metric-affine variational framework in which the
metric and affine connection are the only fundamental variational
fields. General covariance and pseudo-Hermitian invariance constrain the
geometric setting, while the irreducible torsion decomposition resolves
it into contributions with distinct algebraic and differential roles.
Gravitational source-like, constitutive, Maxwell--Proca-type dynamical,
and conservation structures thereby arise at different levels of the
same system rather than from independently introduced fields.

All field-strength-type, response-like, and source-like quantities used
in this interpretation are consequently composite expressions in the
metric and the affine connection. The Maxwell and Proca terminology
refers to the mathematical structure of the reduced equations. In
particular, the transformation of the torsion trace that leaves the
field-strength-type curvature trace invariant describes the gauge
freedom of the reduced Maxwell-type structure; it does not establish an
off-shell gauge symmetry of the complete metric-affine action.

The common organization is the comparison of the full, symmetrized,
and Levi--Civita connections: curvature differences and changes of
covariant divergence generate the non-Riemannian terms in the same
$P_{ij}$. This is an exact geometric decomposition, not an identification
of its separate terms with physical matter.

The present results concern classical equations and geometric
identities. They do not establish which modes propagate or whether
their kinetic energies have the required signs. Results about
instabilities in related nonsymmetric theories
\cite{damour_deser_mccarthy1993} cannot settle those questions for this
action. They require a separate linearization and constraint analysis,
for example around a symmetric Ricci-flat background.

\section*{Acknowledgments}

The author would like to thank Roman Popovych and Michal Marvan for
helpful comments and discussions.

The research was supported by RVO funding for IC47813059.

\bibliographystyle{unsrt}
\bibliography{references}

@article{goenner2004,
  author  = {Goenner, Hubert F. M.},
  title   = {On the History of Unified Field Theories},
  journal = {Living Rev. Relativ.},
  volume  = {7},
  number  = {2},
  year    = {2004},
  doi     = {10.12942/lrr-2004-2}
}

@article{goenner2014,
  author  = {Goenner, Hubert F. M.},
  title   = {On the History of Unified Field Theories. {Part II}. (ca. 1930--ca. 1965)},
  journal = {Living Reviews in Relativity},
  volume  = {17},
  number  = {5},
  year    = {2014},
  doi     = {10.12942/lrr-2014-5}
}

@book{hlavaty1957,
  author    = {Hlavat{\'y}, V{\'a}clav},
  title     = {Geometry of Einstein's Unified Field Theory},
  publisher = {P. Noordhoff},
  address   = {Groningen},
  year      = {1957}
}

@book{tonnelat1966,
  author    = {Tonnelat, Marie-Antoinette},
  title     = {Einstein's Theory of Unified Fields},
  publisher = {Gordon and Breach},
  address   = {New York},
  year      = {1966}
}

@article{ferraris_kijowski1982,
  author  = {Ferraris, Marco and Kijowski, Jerzy},
  title   = {Unified Geometric Theory of Electromagnetic and Gravitational Interactions},
  journal = {Gen. Relativ. Gravit.},
  volume  = {14},
  number  = {1},
  pages   = {37--47},
  year    = {1982},
  doi     = {10.1007/BF00756195}
}

@article{hehl_mccrea_mielke_neeman1995,
  author  = {Hehl, Friedrich W. and McCrea, J. Dermott and Mielke, Eckehard W. and Ne'eman, Yuval},
  title   = {Metric-affine gauge theory of gravity: Field equations, {Noether} identities, world spinors, and breaking of dilation invariance},
  journal = {Physics Reports},
  volume  = {258},
  number  = {1--2},
  pages   = {1--171},
  year    = {1995},
  doi     = {10.1016/0370-1573(94)00111-F}
}

@article{shapiro2002,
  author        = {Shapiro, I. L.},
  title         = {Physical Aspects of the Space-Time Torsion},
  journal       = {Phys. Rep.},
  volume        = {357},
  number        = {2},
  pages         = {113--213},
  year          = {2002},
  doi           = {10.1016/S0370-1573(01)00030-8},
  eprint        = {hep-th/0103093},
  archivePrefix = {arXiv}
}

@article{capozziello_stornaiolo2007,
  author  = {Capozziello, Salvatore and Stornaiolo, Cosimo},
  title   = {Torsion Tensor and Its Geometric Interpretation},
  journal = {Ann. Fond. Louis de Broglie},
  volume  = {32},
  number  = {2--3},
  pages   = {195--214},
  year    = {2007}
}

@book{einstein_meaning_relativity,
  author    = {Einstein, Albert},
  title     = {The Meaning of Relativity},
  publisher = {Princeton University Press},
  address   = {Princeton},
  edition   = {5th},
  year      = {1956},
  note      = {Appendix II: The relativistic theory of the non-symmetric field}
}

@book{landau_lifshitz_fields,
  author    = {Landau, L. D. and Lifshitz, E. M.},
  title     = {The Classical Theory of Fields},
  series    = {Course of Theoretical Physics},
  volume    = {2},
  edition   = {4th},
  publisher = {Pergamon Press},
  address   = {Oxford},
  year      = {1975}
}

@phdthesis{stolin1999_dissertation,
  author       = {Stolín, Oldřich},
  title        = {Nové přístupy k unitárním teoriím},
  school       = {Faculty of Science, Masaryk University},
  address      = {Brno},
  year         = {1999},
  type         = {{Ph.D.} dissertation},
  note         = {Supervised by J. Novotný}
}

@article{chu_leung2021_spin_vacuum,
  author        = {Chu, Chong-Sun and Leung, Chun-Hei},
  title         = {Induced Quantized Spin Current in Vacuum},
  journal       = {Phys. Rev. Lett.},
  volume        = {127},
  number        = {11},
  pages         = {111601},
  year          = {2021},
  doi           = {10.1103/PhysRevLett.127.111601},
  eprint        = {2105.00148},
  archivePrefix = {arXiv},
  primaryClass  = {hep-th}
}

@article{huang_matsuo_taya2019_spin_current_vacuum,
  author        = {Huang, Xu-Guang and Matsuo, Mamoru and Taya, Hidetoshi},
  title         = {Spontaneous Generation of Spin Current from the Vacuum by Strong Electric Fields},
  journal       = {Prog. Theor. Exp. Phys.},
  volume        = {2019},
  number        = {11},
  pages         = {113B02},
  year          = {2019},
  doi           = {10.1093/ptep/ptz112},
  eprint        = {1904.07593},
  archivePrefix = {arXiv},
  primaryClass  = {hep-ph}
}

@article{einstein1945_generalization,
  author  = {Einstein, Albert},
  title   = {A Generalization of the Relativistic Theory of Gravitation},
  journal = {Ann. Math.},
  volume  = {46},
  number  = {4},
  pages   = {578--584},
  year    = {1945},
  doi     = {10.2307/1969197}
}

@article{einstein_straus1946,
  author  = {Einstein, Albert and Straus, Ernst G.},
  title   = {A Generalization of the Relativistic Theory of Gravitation. {II}},
  journal = {Ann. Math.},
  volume  = {47},
  number  = {4},
  pages   = {731--741},
  year    = {1946},
  doi     = {10.2307/1969231}
}

@article{einstein1950_bianchi,
  author  = {Einstein, Albert},
  title   = {The {Bianchi} Identities in the Generalized Theory of Gravitation},
  journal = {Can. J. Math.},
  volume  = {2},
  pages   = {120--128},
  year    = {1950},
  doi     = {10.4153/CJM-1950-011-4}
}

@article{mckellar1979,
  author  = {McKellar, R. J.},
  title   = {Asymmetric Connection Treatment of the {Einstein--Maxwell} Field Equations},
  journal = {Phys. Rev. D},
  volume  = {20},
  number  = {2},
  pages   = {356--361},
  year    = {1979},
  doi     = {10.1103/PhysRevD.20.356}
}

@article{damour_deser_mccarthy1993,
  author        = {Damour, T. and Deser, S. and McCarthy, J.},
  title         = {Nonsymmetric Gravity Theories: Inconsistencies and a Cure},
  journal       = {Phys. Rev. D},
  volume        = {47},
  number        = {4},
  pages         = {1541--1556},
  year          = {1993},
  doi           = {10.1103/PhysRevD.47.1541},
  eprint        = {gr-qc/9207003},
  archivePrefix = {arXiv}
}

@article{moffat1995_new,
  author        = {Moffat, J. W.},
  title         = {A New Nonsymmetric Gravitational Theory},
  journal       = {Phys. Lett. B},
  volume        = {355},
  number        = {3--4},
  pages         = {447--452},
  year          = {1995},
  doi           = {10.1016/0370-2693(95)00670-G},
  eprint        = {gr-qc/9411006},
  archivePrefix = {arXiv}
}

@article{legare_moffat1995,
  author        = {Legar{\'e}, J. and Moffat, J. W.},
  title         = {Field Equations and Conservation Laws in the Nonsymmetric Gravitational Theory},
  journal       = {Gen. Relativ. Gravit.},
  volume        = {27},
  number        = {7},
  pages         = {761--775},
  year          = {1995},
  doi           = {10.1007/BF02105322},
  eprint        = {gr-qc/9412009},
  archivePrefix = {arXiv}
}

@article{ragusa2000,
  author  = {Ragusa, S.},
  title   = {Nonsymmetric Unified Field Theory},
  journal = {Phys. Rev. D},
  volume  = {62},
  number  = {2},
  pages   = {024026},
  year    = {2000},
  doi     = {10.1103/PhysRevD.62.024026},
  note    = {Erratum: Phys. Rev. D 64, 029901 (2001)}
}

@article{wanas_nabil_abdelhamid_elabd2025,
  author  = {Wanas, M. I. and Nabil, Samah and Abdelhamid, Nouran E. and ElAbd, Kyrillos},
  title   = {Differential Identities in {Einstein} Nonsymmetric Geometry},
  journal = {Gravitation and Cosmology},
  volume  = {31},
  number  = {2},
  pages   = {185--194},
  year    = {2025},
  doi     = {10.1134/S0202289325700070}
}

@article{wanas_nabil_sherif2026,
  author  = {Wanas, M. I. and Nabil, Samah and Sherif, A. M.},
  title   = {Parameterization of the Generalized Field Theory},
  journal = {Int. J. Geom. Methods Mod. Phys.},
  volume  = {23},
  number  = {10},
  pages   = {2550263},
  year    = {2026},
  doi     = {10.1142/S0219887825502639}
}

@article{yaremenko2026,
  author  = {Yaremenko, Mykola Ivanovich},
  title   = {Torsion-Mediated Unification in {$Y_n$} Space: {Einstein--Maxwell} Field Equations and Geodesic Dynamics},
  journal = {Int. J. Geom. Methods Mod. Phys.},
  volume  = {24},
  number  = {4},
  pages   = {2650166},
  year    = {2027},
  doi     = {10.1142/S0219887826501665},
  note    = {Published online in 2026}
}

\end{document}